\documentclass[preprint,pre,aps,showpacs,floatfix,nofootinbib]{revtex4}

\usepackage{graphicx}

\newcommand{\fig}[1]{Fig.~\ref{#1}}

\newcommand{\bd}{{\rm b}}

\begin{document}

\title{Effects of the chemomechanical stepping cycle on the
  traffic of molecular motors}

\author{Stefan Klumpp$^{(1)}$, Yan Chai$^{(2)}$, and Reinhard
  Lipowsky$^{(2)}$}

\address{$^{(1)}$Center for Theoretical Biological Physics and
  Department of Physics, University of California at San Diego, 9500
  Gilman Drive, La Jolla, CA 92093-0374 \\ $^{(2)}$Max Planck
  Institute of Colloids and Interfaces, Science Park Golm, 14424
  Potsdam, Germany}

\begin{abstract}
  We discuss effects of the stepping kinetics of molecular motors on
  their traffic behavior on crowded filaments using a simple two-state
  chemomechanical cycle. While the general traffic behavior is quite
  robust with respect to the detailed kinetics of the step, a few
  observable parameters exhibit a strong dependence on these parameters. 
  Most strikingly,
  the effective unbinding rate of the motors may both increase and
  decrease with increasing traffic density, depending on the details
  of the motor step. Likewise the run length either exhibits a strong
  decrease or almost no dependence on the traffic density. We discuss
  recent experiments in light of this analysis.
\end{abstract}


\maketitle


\section{Introduction}

Molecular motors power the transport of various kinds of cargoes
within cells by directed stepping movements along filaments of the
cytoskeleton \cite{Howard2001}. The similarities (and differences)
compared to highway traffic \cite{Traffic} and the fact that cells are
crowded environments \cite{Goodsell1993,Medalia__Baumeister2002} have
stimulated extensive research of traffic phenomena in molecular motor
systems
\cite{Lipowsky__Nieuwenhuizen2001,Klumpp_Lipowsky2003,Kruse_Sekimoto2002,Parmeggiani__Frey2003,Evans__Santen2003,Klumpp_Lipowsky2004,Klumpp_Lipowsky_PRE2004,Klein__Juelicher2005,Nishinari__Chowdhury2005}.

Most studies of molecular motor traffic have modeled the stochastic
stepping of an individual motor by a single Poissonian step, which is
modified in dense traffic by an exclusion rule similar to the
well-studied asymmetric simple exclusion process (ASEP)
\cite{MacDonald__Pipkin1968,Krug1991,Derrida__Pasquier1993,Schuetz_Domany1993}:
If the site to which a motor attempts to step is not accessible,
because another motor (or any other kind of obstacle) is bound there,
the step is rejected. Stepping of a molecular motor is however a complex
process and consists of a series of transitions between different motor
states and corresponding conformational changes \cite{Liepelt_Lipowsky2007}.
Several recent studies have therefore incorporated detailed kinetic
models of the steps into models for molecular motor traffic
\cite{Nishinari__Chowdhury2005,Wang__Wang2005,Greulich__Chowdhury2007}.
However, the main feature of molecular motor traffic visible in
experiments
\cite{Leduc__Prost2004,Konzack__Fischer2005,Nishinari__Chowdhury2005},
the emergence of traffic jam-like density profiles, with a region of
high density of bound motors separated from a low-density region by a
sharp interface
\cite{Lipowsky__Nieuwenhuizen2001,Klumpp__Lipowsky2005,Parmeggiani__Frey2003},
as well as the basic features of the phase diagrams for transport in
open systems \cite{Klumpp_Lipowsky2003,Parmeggiani__Frey2003} are very
robust with respect to such extensions of the simplest models.

In this paper, we discuss observable effects of the stepping
kinetics on the traffic behavior. To be explicit we consider the
simplest possible stepping cycle, which consists of transitions
between two motor conformations or internal states of the motor, 
with only one transition involving
movement of the motor to the next binding site on the filament 
as shown in \fig{model}. A few generalizations for models that incorporate reverse transitions and backward steps are briefly discussed in the appendix.

\section{Model without motor binding/unbinding}

\begin{figure}[htbp]
  \centering
  \includegraphics[angle=0,width=.9\textwidth]{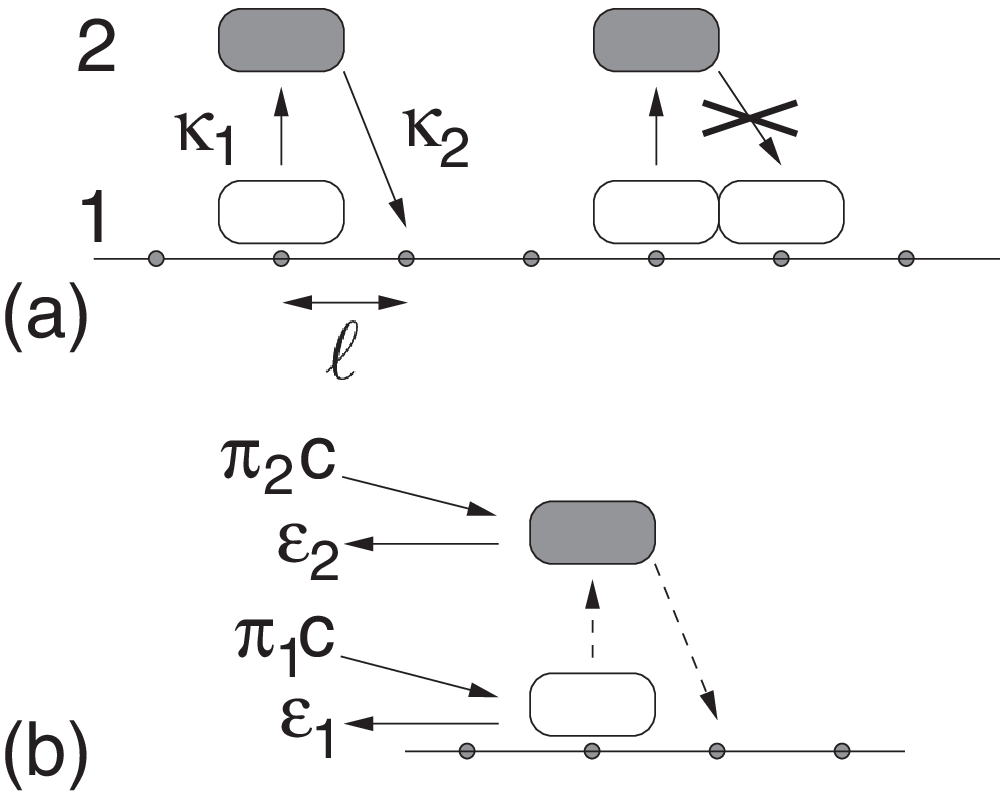}
  \caption{Model for traffic of molecular motors with a two-state chemomechanical cycle: (a) Motors can be in two conformational states, state 1 (white) and 2 (grey). The two states may be considered as a weakly bound state and a strongly bound state. A motor in state 1 can make a transition to state 2 with rate $\kappa_1$, while remaining at the same site on the filament. A motor in state 2 makes a transition to state 1 with rate $\kappa_2$. The latter transition is accompanied by a step of size $\ell$ to the next site on the filament and occurs only if the forward neighbor site is not occupied by another motor in either state 1 or 2. (b) Motors in state 1 and 2 unbind from the filament with rates $\epsilon_1$ and $\epsilon_2$, respectively. Likewise, a free binding site on the filament becomes occupied by a motor in state 1 or 2 with rates $\pi_1 c$ and $\pi_2 c$, respectively, which depend on the concentration $c$ of unbound motors. Unbound motors may not bind to sites already occupied by a motor in either state.}
  \label{model}
\end{figure}

We first consider the case without exchange of motors with a
reservoir, i.e. we neglect that fact that bound motors unbind from
their filamentous track and unbound motors bind to it. In the simplest 
model, the motor cycles between two conformational states, state 1 and 2, which may be considered as a weakly bound and a strongly bound state as in a kinetic scheme for kinesin 1 \cite{Seitz_Surrey2006}. In that case,
the model with two substeps is defined by two rates, $\kappa_1$ and $\kappa_2$.
The first one describes the transitions from state 1 to state 2 with
the motor position along the filament unchanged, while the second
describes the actual step to the next site, which is accompanied by a
transition from motor state 2 (the strongly bound state of kinesin) to state 1 (the weakly bound state) and has step size $\ell$, as depicted in
\fig{model}(a). Binding and unbinding of motors as indicated in \fig{model}(b) 
will be incorporated into the model in the next section. Throughout this paper we use periodic boundary conditions; we note however that more realistic open boundaries will lead to the same results if the system size is long compared to the run length of the motors, i.e. the distance a motor moves along the filament before unbinding \cite{Traffic}. Under the latter condition, which is typically fulfilled for cytoskeletal motors with run lengths of $\sim$ 1$\mu$m and filaments with typical lengths of tens of microns \cite{Howard2001}, the choice of the boundary conditions is only reflected in boundary layers of a size comparable to the run length \cite{Traffic}. 

\begin{figure}[htbp]
  \centering
  \includegraphics[angle=0,width=.9\textwidth]{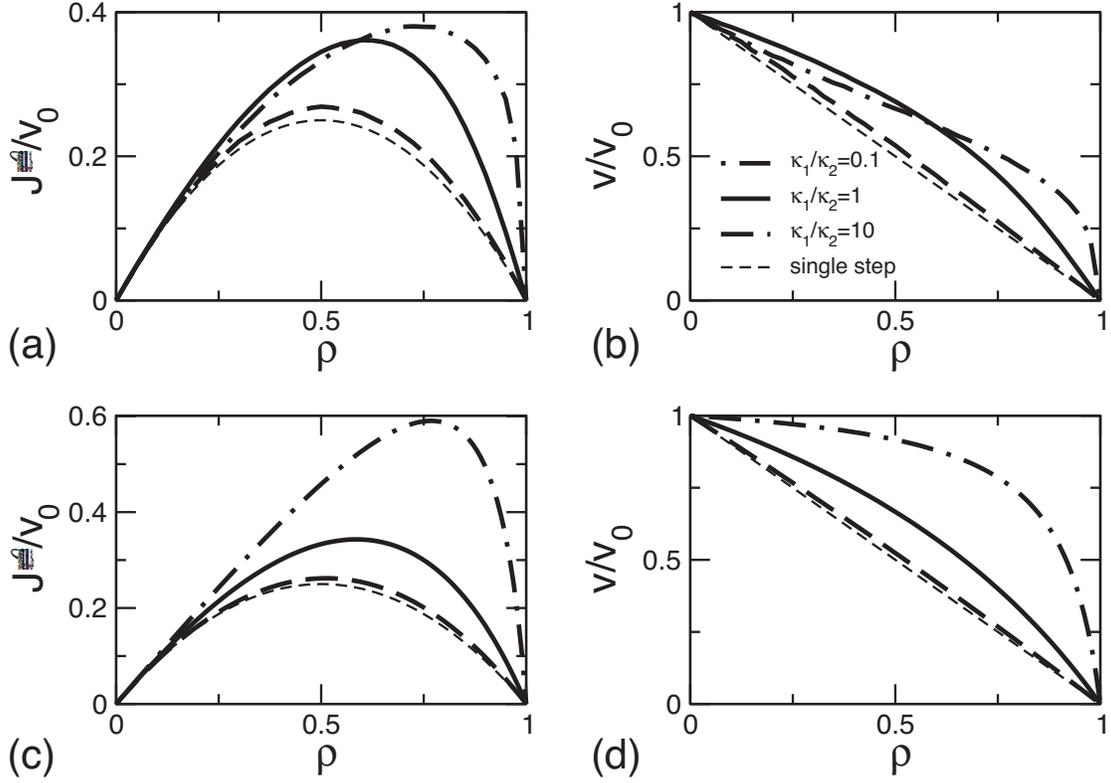}
  \caption{Effect of the chemomechanical stepping cycle: (a,c) Motor current
    $J$ and (b,d) velocity $v$ as functions of the motor density
    $\rho$ on the filament for different choices of the rates $\kappa_1$
    and $\kappa_2$. (a) and (b) show results from simulations, while (c)
    and (d) show the corresponding mean field results. All plots shown here have been obtained for the simplified models in which the motors do not unbind from the filament. Both current and velocity have been normalized
    by the single motor velocity $v_0$, which is taken to be an experimentally determined quantity. 
    With this normalization these quantities are functions of the ratio of $\kappa_1/\kappa_2$ only.}
  \label{J_V}
\end{figure}

\fig{J_V}(a) shows the motor current as a function of the motor
density $\rho$ on the filament as obtained from simulations.  Here and in the
following plots, we normalize the motor current $J$ and the motor velocity $v$ by the single motor
velocity $v_0=\kappa_1\kappa_2\ell/(\kappa_1+\kappa_2)$, which we consider as a known quantity given by experimental data.  In contrast to the quantities $J$, $v$ and $v_0$, the normalized current $J\ell/v_0$ and the normalized velocity $v/v_0$, which will be discussed below, depend only on the ratio $\kappa_1/\kappa_2$ of the two transition rates and not on the absolute values  of $\kappa_1$ and $\kappa_2$. Therefore $J\ell/v_0$ was plotted for different choices of  $\kappa_1/\kappa_2$ in \fig{J_V}(a).  In the limit $\kappa_1\gg \kappa_2$, the
actual step is rate-limiting and the transition from state 1 to state 2, 
which does not comprise movement
along the filament, can be neglected. In this limit, the two-step
model becomes equivalent to the usual ASEP and the current is given by
$J=(v_0/\ell)\rho(1-\rho)$, symmetric around $\rho=1/2$. If however the
conformational transition cannot be neglected, this symmetry is lost
and the maximum of the current is shifted towards higher densities. At
the same time, the value of the maximal current increases. The absence of particle-hole symmetry, i.e.\ symmetry upon interchanging $\rho$ and $(1-\rho)$, can be understood by considering the velocity of rare individual particles or holes, i.e.\ the limits $\rho\approx 0$ and $\rho\approx 1$, respectively. While a single motor particle moves with velocity $v_0$, a single hole in a fully occupied lattice moves with the hole velocity $v_H=\kappa_2\ell$, as all particles it encounters had sufficient time for the transition to state 2. This velocity is always larger than the particle velocity $v_0$, as $v_0/v_H=\kappa_1/(\kappa_1+\kappa_2)<1$. Thus the current, which is given by $J= \kappa_2(1-\rho)$ for $\rho\approx 1$, drops more steeply for $\rho\approx 1$ than it increases for small $\rho$.

While the current is the most important characteristics of a traffic
system from a theoretical point of view, it is not easily accessible
in experiments. A quantity that can be measured directly is the motor
velocity, which can be determined by tracking individual labeled motors
in a background of unlabeled motors \cite{Seitz_Surrey2006}. The
velocity $v$ can be expressed in terms of the current $J$ using the
relation $v=J\ell/\rho$. The density-dependence of the motor velocity is
shown in \fig{J_V}(b). In contrast to the usual ASEP, which is obtained in the
limit $\kappa_1\gg \kappa_2$ and for which the velocity exhibits a linear decrease as a
function of the density $\rho$, the velocity is a convex function of
the density in the two-step model. The reduction of the motor velocity
is thus less pronounced at low motor densities, in particular if the
actual step is not rate-limiting, i.e. if $\kappa_2\gg \kappa_1$. This means that
a strong slowing down of the motors due to the traffic may only be
observed for very dense motor traffic. At intermediate densities, the effect of exclusion is reduced for large $\kappa_2$. This can be interpreted as a partially synchronized motion of motors: If one or several motors in state 2 are directly behind a motor in state 1, they follow almost immediately when the leading motor steps forward.   

The density-dependence of the motor current and the motor velocity can
be described analytically using a mean-field approximation: all
traffic effects can be subsumed into an effective rate for the actual
stepping by replacing the rate $\kappa_2$ by $\kappa_2(1-\rho)$. Then the
probabilities $P_1$ and $P_2$ that the motor is in conformation 1 or 2
are calculated as for a single motor. This approximation leads to
\begin{equation}
  P_1= \frac{\kappa_2(1-\rho)}{\kappa_1+\kappa_2(1-\rho)}\quad{\rm and}\quad P_2=\frac{\kappa_1}{\kappa_1+\kappa_2(1-\rho)},
\end{equation}
and thus to
\begin{equation}
  J(\rho) = \rho \kappa_2 P_2=\frac{\kappa_1 \kappa_2\rho(1-\rho)}{\kappa_1+\kappa_2(1-\rho)}
\end{equation}
and
\begin{equation}
  v(\rho)  = \kappa_2\ell P_2 =\frac{\kappa_1 \kappa_2(1-\rho)\ell}{\kappa_1+\kappa_2(1-\rho)}. 
\end{equation}
These expressions are plotted in \fig{J_V}(c) and (d), again normalized by the single motor
velocity $v_0$. These plots show qualitatively the same behavior as
the simulation data in \fig{J_V}(a) and (b).  
However, except for the limiting case $\kappa_1 \gg \kappa_2$,
which corresponds to the usual ASEP, the mean field approximation
overestimates the effects of the conformational transition. In
particular, in the limit $\kappa_2\gg \kappa_1$ it predicts a linear increase of
the current, $J=(v_0/\ell)\rho$, and a constant velocity, $v=v_0$, up to almost the maximal density $\rho=1$.
The maximal current thus approaches $J\ell/v_0=1$ and
occurs at a density value that approaches $\rho=1$. In contrast, the
density-dependence of the current obtained from simulations shows
pronounced deviations from the linear density dependence of the current
even for low $\kappa_1/\kappa_2$ and appears to converge to a function with a
maximum of $J/(v_0/\ell)\approx 0.37$ at $\rho\approx 0.75$. The curves for
$\kappa_1/\kappa_2=0.1$ in \fig{J_V} are close to this limit.

\section{Model with binding/unbinding kinetics}

Next, we consider the effect of motor binding to and unbinding from
the filament [\fig{model}(b)]. Even the simplest model with only two internal states 
of the motor allows many different implementations of binding and unbinding:
unbinding may occur from state 1 and/or state 2, described by the
rates $\epsilon_1$ and $\epsilon_2$, and unbound motors may also
become bound in either state, described by the rates $\pi_{1}c$ and
$\pi_2c$, where $\pi_{1,2}$ are the second-order binding rates and 
$c$ is the concentration of unbound motors. For simplicity, we 
restrict the discussion to the two cases,
where unbinding occurs predominantly in one of the two states.
Furthermore, our simulations indicate that whether newly bound motors
are in state 1 or state 2 has only a small effect as long as the
motors are processive, i.e. if they typically make many steps before
unbinding \cite{en1}.

\begin{figure}[htbp]
  \centering
  \includegraphics[angle=0,width=.9\textwidth]{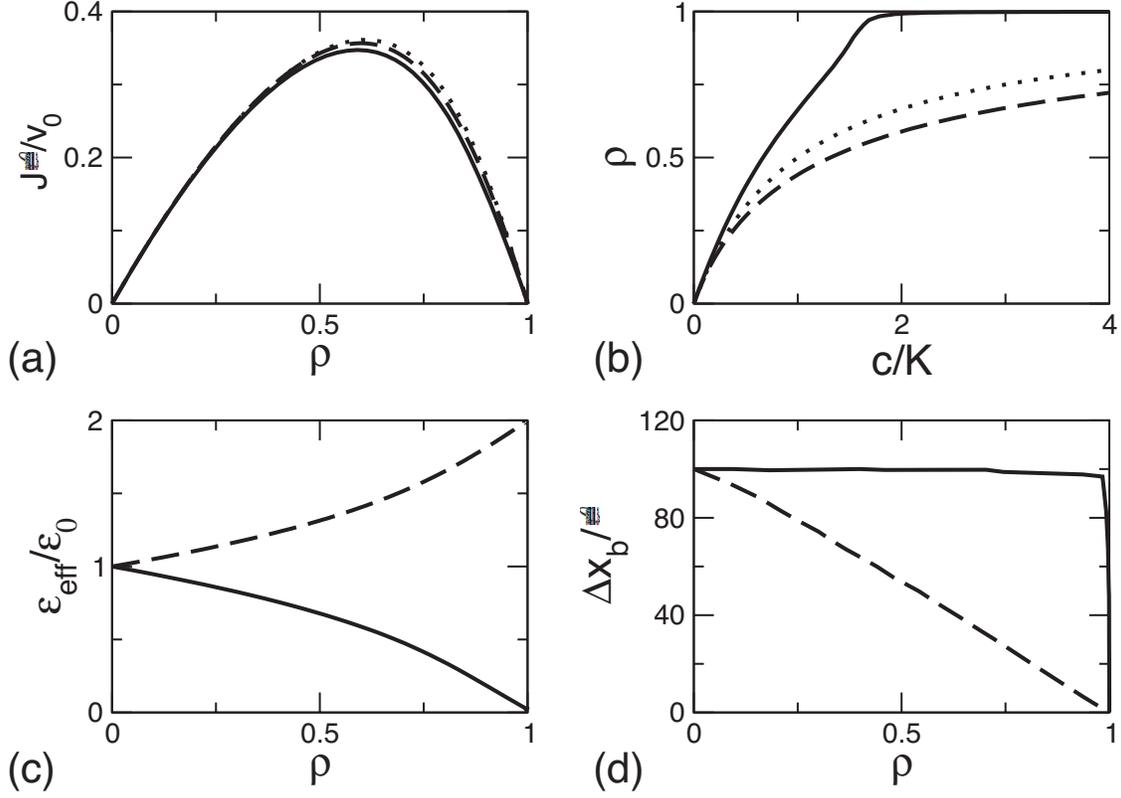}
  \caption{Effect of motor binding and unbinding: (a) Motor current
    $J$ as a function of the motor density $\rho$, (b) motor density
    $\rho$ as a function of the unbound motor concentration $c$,
    normalized by the dissociation constant
    $K=\epsilon_0/(\pi_1+\pi_2)$, (c) effective unbinding rate
    $\epsilon_{\rm eff}$ and (d) average run length $\Delta x_\bd$ as
    functions of the bound motor density $\rho$. All plots show
    simulation results for models with unbinding predominantly from
    state 1 (solid lines, $\epsilon_1=\kappa_1/100$, $\epsilon_2=\epsilon_1/100$,
    $\kappa_1=\kappa_2$) and with unbinding from state 2 (dashed lines, $\epsilon_1=0$, 
    $\epsilon_2=\kappa_2/100$, $\kappa_1=\kappa_2$). The dotted line
    in (a) shows the corresponding curve without binding/unbinding,
    the dotted line in (b) the Langmuir-type binding curve for the
    same single motor unbinding rate $\epsilon_0$.}
  \label{unbind}
\end{figure}

Results from simulations that include binding and unbinding of motors
are shown in \fig{unbind}.  The solid and dashed curves in \fig{unbind}
correspond to cases where unbinding occurs predominantly from state 1,
with $\epsilon_1=\kappa_1/100$ and $\epsilon_2=\epsilon_1/100$ \cite{en2},
and
only from state 2, with $\epsilon_2=\kappa_2/100$ and $\epsilon_1=0$, 
respectively.  In both
cases, the unbinding rates are chosen to match a run length of 100
steps for the single motor.  For processive motors, including motor
binding and unbinding has only a small effect on the
density-dependence of the motor current [\fig{unbind} (a)] and the
motor velocity compared to the case without binding/unbinding (dotted
line).  The density $\rho$ of bound motors, however, is now determined
by the binding/unbinding equilibrium.  \fig{unbind}(b) shows the bound
motor density as a function of the unbound motor concentration $c$, which is normalized by the desorption constant $K\equiv \epsilon_{0}/(\pi_1+\pi_2)$. Here $\epsilon_0$ is the effective unbinding rate of a single motor,  which is obtained from the general effective unbinding rate $\epsilon_{\rm eff}=\epsilon_1 P_1+\epsilon_2P_2$ by taking the limit $\rho=0$, which leads to $\epsilon_0=(\epsilon_1\kappa_2+\epsilon_2\kappa_1)/(\kappa_1+\kappa_2)$.  These curves
are compared to simple Langmuir-type binding with the same effective
unbinding rate $\epsilon_{0}$ for
a single motor (dotted line), as used in models without internal
states.
While all three curves are the same for small binding rates (or
unbound motor concentrations), there are marked differences for high
binding rates. If unbinding occurs from state 1, binding is clearly
stronger than in simple Langmuir-type binding \cite{en3}, while it is weaker if
unbinding occurs from state 2. This behavior is not surprising,
because, in denser traffic, motors spend a larger fraction of time in
state 2, which increases the effective unbinding rate $\epsilon_{\rm eff}$ if
unbinding occurs from state 2, but decreases it if
unbinding occurs from state 1, as shown in \fig{unbind}(c).

Closely connected with the unbinding rate is the run length or 
processivity, i.e. the
number of steps a motor takes before detaching from the filament,
which is shown in \fig{unbind}(d) as a function of the motor density
on the filament. For simple Poissonian steppers, the unbinding rate is
independent of the motor density, but the run length decreases
linearly with increasing density due to the linear decrease of the
velocity. For motors with a conformational stepping cycle, the run
length also decreases if unbinding occurs from state 2 [although the decrease
is not necessarily linear, in particular if $\kappa_1\gg \kappa_2$, due
to the convex density-dependence of the velocity as shown in
\fig{J_V}(b)]. If unbinding occurs however predominantly from state 1,
the run length remains unaffected by the density of the motor traffic
up to very high densities (where the small unbinding rate from state 2
becomes dominant). In this case, unbinding is characterized by a
constant unbinding probability \emph{per step} rather than a constant
unbinding rate as in the case of Poissonian steppers.

\section{Experimental relevance}

We have discussed the effects of the chemomechanical stepping cycle
of a molecular motor on the traffic of many molecular motors using a
simple two-state chemical cycle. For this model we have determined
observable quantities, in particular the motor velocity and the
average run length, which can be measured by tracking individual
motors on a crowded filament \cite{Seitz_Surrey2006}.

Our analysis shows that the more complex stepping kinetics tend to
diminish the slowing down of motors due to traffic congestion, but this effect is not very large. 
It is possible that this
effect contributes to the surprisingly small decrease of the velocity
observed in the experiment of Seitz and Surrey
\cite{Seitz_Surrey2006}. In this experiment, the velocity of kinesin 1 
motors remained almost constant up to unbound motor concentrations that resulted in nominal bound motor densities of $\approx$0.3-0.5 and reduced the
motor binding rate about two-fold. The interpretation of this
experiment is however complicated by another observation showing that
microtubules saturated with motors exhibit an approximately two-fold
higher motor density in the presence of a non-hydrolysable ATP
analogous than in the presence of ATP. It is thus rather likely that the
specific geometry of binding of motors to the filament also plays an
important role, in particular whether they bind with one or two heads and 
how many sites a motor occupies on the filament \cite{ProcKanpur}. A
larger size of the particles also diminishes the decrease of the
velocity \cite{MacDonald__Pipkin1968,ProcKanpur}, so one may
speculate that the relatively constant velocity observed by Seitz and
Surrey is due to a combination of (at least) two weak effects.

A more pronounced effect of the cycle is seen in the
density-dependence of the run length and the effective unbinding rate.
Depending on the state from which unbinding from the filament
predominantly occurs, the effective unbinding rate may both increase or
decrease with increasing traffic density, while the run length
decreases or remains essentially constant. The latter behavior has
been observed in the experiment of ref.\ \cite{Seitz_Surrey2006}. In
this experiment, the time a motor spends bound to the filament before
unbinding was found to increase about two-fold over the studied
range of motor densities, indicating that the unbinding rate decreases
in dense traffic, while the run length remained approximately
constant. Another study \cite{Crevel__Cross2004} however reported the
exact opposite, namely that unbinding is increased due to crowding.
The reason for the discrepancy is not known, but there are several
important differences in the experimental conditions. In the
experiment of ref.\ \cite{Crevel__Cross2004}, microtubules were covered
by very high densities of kinesin (up to one kinesin dimer per
microtubule binding site) in the absence of nucleotides and unbinding
was measured indirectly by light scattering during the transient
dynamics after the addition of ATP. In contrast, the experiment of
ref.\ \cite{Seitz_Surrey2006} used lower motor densities, addressed the
steady-state of the motor traffic, and motor properties were measured
directly by tracking individual motors. The observation that different
amounts of kinesin are needed to saturate microtubules in the presence
of different nucleotides (ATP versus non-hydrolysable analogue)
\cite{Seitz_Surrey2006} also suggests that kinesins may be bound in
different ways in the two experiments. In general, even the equilibrium binding of
dimeric motors to filaments (in the absence of ATP, i.e.\ without active movements) 
may exhibit a range of stoichiometries and rather complex dynamics \cite{Vilfan_Frey2002}.

Finally, the main effect of crowded molecular motor traffic is the
emergence of a traffic-jam-like domain of high motor density at the
end of a filament
\cite{Lipowsky__Nieuwenhuizen2001,Klumpp__Lipowsky2005,Parmeggiani__Frey2003},
as observed in several experiments
\cite{Leduc__Prost2004,Konzack__Fischer2005,Nishinari__Chowdhury2005}.
If the filament is in contact with a large reservoir of unbound
motors, the length of the high density domain or traffic jam is of the
order of the run length in ASEP-type models \cite{Traffic}. If the
unbinding rate is reduced in dense traffic, as in our model with
unbinding from state 1 and as suggested by the experiment of ref.
\cite{Seitz_Surrey2006}, one may expect to observe much longer
traffic jams.  Extending the arguments given in ref.\ \cite{Traffic}, one obtains the estimate
that the jam length is larger than the run length by a
factor $\epsilon_0/ \epsilon (\rho)$.

\section*{Appendix}

The model we have discussed in this article represents the simplest possible chemomechanical cycle of a molecular motor with two subsequent transitions that are treated as irreversible. This minimal model of a stepping cycle provides a good description of the dynamics of kinesin motors under typical experimental conditions (low concentrations of the hydrolysis products ADP and P$_{\rm i}$, no opposing force), see, e.g., Ref.\ \cite{Seitz_Surrey2006}, where backward steps are rare \cite{Carter_Cross2005,Liepelt_Lipowsky2007}, but should not be expected to be valid under all possible experimental conditions or for other types of motors such as dyneins which exhibit backward steps more frequently than kinesins \cite{Ross__Holzbaur2006,Gennerich__Vale2007}. On the other hand, it is straightforward to include backward steps in the theoretical framework used here as already described in Ref.\ \cite{Lipowsky__Nieuwenhuizen2001}. To test to what extent our results remain valid if the dynamics of motor stepping are more complex, we simulated several variants of our model that include reverse transitions and backward steps. All these variants exhibit qualitatively the same behavior as the simple model. 

Surprisingly, the main difference to the simple model is already obtained when the reverse transition for the transition from state 1 to state 2 is included: In the original model with irreversible transitions, a motor in dense traffic often waits in state 2 until the site in front becomes available. If the reverse transition is included such a motor will switch between states 1 and 2 while waiting, so the probability that the motor is in state 2 is reduced in dense traffic compared to the irreversible model. As a consequence the motor velocity is slightly reduced and the distinction between the cases with unbinding from state 1 or 2 becomes less pronounced if the reverse rate is sufficiently large. In particular, the run length exhibits a weak decrease with increasing motor density (rather than being constant) in the case where unbinding occurs mainly from state 1. Essentially the same behavior was observed for a cycle that included reverse transition for both the internal transition and the actual step. 

Finally we also simulated the case where a motor in state 2 may make either a forward step or (with a smaller rate) a backward step, which provides the simplest model where forward and backward steps occur along different reaction pathways, a situation suggested for kinesin by both experiments and modeling \cite{Carter_Cross2005,Liepelt_Lipowsky2007}. Our simulation results for this case strongly resemble those for the simplest cycle without backsteps or reverse transitions.

\begin{acknowledgments}
The authors thank A. Seitz and T. Surrey for discussions of their experiments. SK was supported by Deutsche Forschungsgemeinschaft (grants KL818/1-1 and 1-2) and by the National Science Foundation through the Center for Theoretical Biological Physics (grants PHY-0216576 and 0225630).
\end{acknowledgments}



\newpage

\begin{thebibliography}{25}
\expandafter\ifx\csname natexlab\endcsname\relax\def\natexlab#1{#1}\fi
\expandafter\ifx\csname bibnamefont\endcsname\relax
  \def\bibnamefont#1{#1}\fi
\expandafter\ifx\csname bibfnamefont\endcsname\relax
  \def\bibfnamefont#1{#1}\fi
\expandafter\ifx\csname citenamefont\endcsname\relax
  \def\citenamefont#1{#1}\fi
\expandafter\ifx\csname url\endcsname\relax
  \def\url#1{\texttt{#1}}\fi
\expandafter\ifx\csname urlprefix\endcsname\relax\def\urlprefix{URL }\fi
\providecommand{\bibinfo}[2]{#2}
\providecommand{\eprint}[2][]{\url{#2}}

\bibitem[{\citenamefont{Howard}(2001)}]{Howard2001}
\bibinfo{author}{\bibfnamefont{J.}~\bibnamefont{Howard}},
  \emph{\bibinfo{title}{Mechanics of Motor Proteins and the Cytoskeleton}}
  (\bibinfo{publisher}{Sinauer Associates}, \bibinfo{address}{Sunderland
  (Mass.)}, \bibinfo{year}{2001}).

\bibitem[{\citenamefont{Klumpp et~al.}(2007)\citenamefont{Klumpp, M{\"u}ller,
  and Lipowsky}}]{Traffic}
\bibinfo{author}{\bibfnamefont{S.}~\bibnamefont{Klumpp}},
  \bibinfo{author}{\bibfnamefont{M.~J.~I.} \bibnamefont{M{\"u}ller}},
  \bibnamefont{and} \bibinfo{author}{\bibfnamefont{R.}~\bibnamefont{Lipowsky}},
  in \emph{\bibinfo{booktitle}{Traffic and Granular Flow '05}}, edited by
  \bibinfo{editor}{\bibfnamefont{A.}~\bibnamefont{Schadschneider}},
  \bibinfo{editor}{\bibfnamefont{T.}~\bibnamefont{{P\"oschel}}},
  \bibinfo{editor}{\bibfnamefont{R.}~\bibnamefont{{K\"uhne}}},
  \bibinfo{editor}{\bibfnamefont{M.}~\bibnamefont{Schreckenberg}},
  \bibnamefont{and} \bibinfo{editor}{\bibfnamefont{D.~E.} \bibnamefont{Wolf}}
  (\bibinfo{publisher}{Springer}, \bibinfo{address}{Berlin},
  \bibinfo{year}{2007}), pp. \bibinfo{pages}{251--261}.

\bibitem[{\citenamefont{Goodsell}(1993)}]{Goodsell1993}
\bibinfo{author}{\bibfnamefont{D.~S.} \bibnamefont{Goodsell}},
  \emph{\bibinfo{title}{The Machinery of Life}} (\bibinfo{publisher}{Springer},
  \bibinfo{address}{New York}, \bibinfo{year}{1993}).

\bibitem[{\citenamefont{Medalia et~al.}(2002)\citenamefont{Medalia, Weber,
  Frangakis, Nicastro, Gehrisch, and Baumeister}}]{Medalia__Baumeister2002}
\bibinfo{author}{\bibfnamefont{O.}~\bibnamefont{Medalia}},
  \bibinfo{author}{\bibfnamefont{I.}~\bibnamefont{Weber}},
  \bibinfo{author}{\bibfnamefont{A.~S.} \bibnamefont{Frangakis}},
  \bibinfo{author}{\bibfnamefont{D.}~\bibnamefont{Nicastro}},
  \bibinfo{author}{\bibfnamefont{G.}~\bibnamefont{Gehrisch}}, \bibnamefont{and}
  \bibinfo{author}{\bibfnamefont{W.}~\bibnamefont{Baumeister}},
  \bibinfo{journal}{Science} \textbf{\bibinfo{volume}{298}},
  \bibinfo{pages}{1209} (\bibinfo{year}{2002}).

\bibitem[{\citenamefont{Lipowsky et~al.}(2001)\citenamefont{Lipowsky, Klumpp,
  and Nieuwenhuizen}}]{Lipowsky__Nieuwenhuizen2001}
\bibinfo{author}{\bibfnamefont{R.}~\bibnamefont{Lipowsky}},
  \bibinfo{author}{\bibfnamefont{S.}~\bibnamefont{Klumpp}}, \bibnamefont{and}
  \bibinfo{author}{\bibfnamefont{T.~M.} \bibnamefont{Nieuwenhuizen}},
  \bibinfo{journal}{Phys.\ Rev.\ Lett.} \textbf{\bibinfo{volume}{87}},
  \bibinfo{pages}{108101} (\bibinfo{year}{2001}).

\bibitem[{\citenamefont{Klumpp and Lipowsky}(2003)}]{Klumpp_Lipowsky2003}
\bibinfo{author}{\bibfnamefont{S.}~\bibnamefont{Klumpp}} \bibnamefont{and}
  \bibinfo{author}{\bibfnamefont{R.}~\bibnamefont{Lipowsky}},
  \bibinfo{journal}{J.\ Stat.\ Phys.} \textbf{\bibinfo{volume}{113}},
  \bibinfo{pages}{233} (\bibinfo{year}{2003}).

\bibitem[{\citenamefont{Kruse and Sekimoto}(2002)}]{Kruse_Sekimoto2002}
\bibinfo{author}{\bibfnamefont{K.}~\bibnamefont{Kruse}} \bibnamefont{and}
  \bibinfo{author}{\bibfnamefont{K.}~\bibnamefont{Sekimoto}},
  \bibinfo{journal}{Phys.\ Rev.\ E} \textbf{\bibinfo{volume}{66}},
  \bibinfo{pages}{031904} (\bibinfo{year}{2002}).


\bibitem[{\citenamefont{Parmeggiani et~al.}(2003)\citenamefont{Parmeggiani,
  Franosch, and Frey}}]{Parmeggiani__Frey2003}
\bibinfo{author}{\bibfnamefont{A.}~\bibnamefont{Parmeggiani}},
  \bibinfo{author}{\bibfnamefont{T.}~\bibnamefont{Franosch}}, \bibnamefont{and}
  \bibinfo{author}{\bibfnamefont{E.}~\bibnamefont{Frey}},
  \bibinfo{journal}{Phys.\ Rev.\ Lett.} \textbf{\bibinfo{volume}{90}},
  \bibinfo{pages}{086601} (\bibinfo{year}{2003}).

\bibitem[{\citenamefont{Evans et~al.}(2003)\citenamefont{Evans, Juh{\'a}sz, and
  Santen}}]{Evans__Santen2003}
\bibinfo{author}{\bibfnamefont{M.~R.} \bibnamefont{Evans}},
  \bibinfo{author}{\bibfnamefont{R.}~\bibnamefont{Juh{\'a}sz}},
  \bibnamefont{and} \bibinfo{author}{\bibfnamefont{L.}~\bibnamefont{Santen}},
  \bibinfo{journal}{Phys.\ Rev.\ E} \textbf{\bibinfo{volume}{68}},
  \bibinfo{pages}{026117} (\bibinfo{year}{2003}).

\bibitem[{\citenamefont{Klumpp and
  Lipowsky}(2004{\natexlab{a}})}]{Klumpp_Lipowsky2004}
\bibinfo{author}{\bibfnamefont{S.}~\bibnamefont{Klumpp}} \bibnamefont{and}
  \bibinfo{author}{\bibfnamefont{R.}~\bibnamefont{Lipowsky}},
  \bibinfo{journal}{Europhys.\ Lett.} \textbf{\bibinfo{volume}{66}},
  \bibinfo{pages}{90} (\bibinfo{year}{2004}{\natexlab{a}}).

\bibitem[{\citenamefont{Klumpp and
  Lipowsky}(2004{\natexlab{b}})}]{Klumpp_Lipowsky_PRE2004}
\bibinfo{author}{\bibfnamefont{S.}~\bibnamefont{Klumpp}} \bibnamefont{and}
  \bibinfo{author}{\bibfnamefont{R.}~\bibnamefont{Lipowsky}},
  \bibinfo{journal}{Phys.\ Rev.\ E} \textbf{\bibinfo{volume}{70}},
  \bibinfo{pages}{066104} (\bibinfo{year}{2004}{\natexlab{b}}).

\bibitem[{\citenamefont{Klein et~al.}(2005)\citenamefont{Klein, Kruse,
  Cuniberti, and J{\"u}licher}}]{Klein__Juelicher2005}
\bibinfo{author}{\bibfnamefont{G.~A.}~\bibnamefont{Klein}},
  \bibinfo{author}{\bibfnamefont{K.}~\bibnamefont{Kruse}},
  \bibinfo{author}{\bibfnamefont{G.}~\bibnamefont{Cuniberti}},
  \bibnamefont{and}
  \bibinfo{author}{\bibfnamefont{F.}~\bibnamefont{J\"ulicher}},
  \bibinfo{journal}{Phys.\ Rev.\ Lett.} \textbf{\bibinfo{volume}{94}},
  \bibinfo{pages}{108102} (\bibinfo{year}{2005}).

\bibitem[{\citenamefont{Nishinari et~al.}(2005)\citenamefont{Nishinari, Okada,
  Schadschneider, and Chowdhury}}]{Nishinari__Chowdhury2005}
\bibinfo{author}{\bibfnamefont{K.}~\bibnamefont{Nishinari}},
  \bibinfo{author}{\bibfnamefont{Y.}~\bibnamefont{Okada}},
  \bibinfo{author}{\bibfnamefont{A.}~\bibnamefont{Schadschneider}},
  \bibnamefont{and}
  \bibinfo{author}{\bibfnamefont{D.}~\bibnamefont{Chowdhury}},
  \bibinfo{journal}{Phys.\ Rev.\ Lett.} \textbf{\bibinfo{volume}{95}},
  \bibinfo{pages}{118101} (\bibinfo{year}{2005}).

\bibitem[{\citenamefont{MacDonald et~al.}(1968)\citenamefont{MacDonald, Gibbs,
  and Pipkin}}]{MacDonald__Pipkin1968}
\bibinfo{author}{\bibfnamefont{C.~T.} \bibnamefont{MacDonald}},
  \bibinfo{author}{\bibfnamefont{J.~H.} \bibnamefont{Gibbs}}, \bibnamefont{and}
  \bibinfo{author}{\bibfnamefont{A.~C.} \bibnamefont{Pipkin}},
  \bibinfo{journal}{Biopolymers} \textbf{\bibinfo{volume}{6}},
  \bibinfo{pages}{1} (\bibinfo{year}{1968}).

\bibitem[{\citenamefont{Krug}(1991)}]{Krug1991}
\bibinfo{author}{\bibfnamefont{J.}~\bibnamefont{Krug}},
  \bibinfo{journal}{Phys.\ Rev.\ Lett.} \textbf{\bibinfo{volume}{67}},
  \bibinfo{pages}{1882} (\bibinfo{year}{1991}).

\bibitem[{\citenamefont{Derrida et~al.}(1993)\citenamefont{Derrida, Evans,
  Hakim, and Pasquier}}]{Derrida__Pasquier1993}
\bibinfo{author}{\bibfnamefont{B.}~\bibnamefont{Derrida}},
  \bibinfo{author}{\bibfnamefont{M.~R.} \bibnamefont{Evans}},
  \bibinfo{author}{\bibfnamefont{V.}~\bibnamefont{Hakim}}, \bibnamefont{and}
  \bibinfo{author}{\bibfnamefont{V.}~\bibnamefont{Pasquier}},
  \bibinfo{journal}{J.\ Phys.\ A:\ Math.\ Gen.} \textbf{\bibinfo{volume}{26}},
  \bibinfo{pages}{1493} (\bibinfo{year}{1993}).

\bibitem[{\citenamefont{Sch\"utz and Domany}(1993)}]{Schuetz_Domany1993}
\bibinfo{author}{\bibfnamefont{G.}~\bibnamefont{Sch\"utz}} \bibnamefont{and}
  \bibinfo{author}{\bibfnamefont{E.}~\bibnamefont{Domany}},
  \bibinfo{journal}{J.\ Stat.\ Phys.} \textbf{\bibinfo{volume}{72}},
  \bibinfo{pages}{277} (\bibinfo{year}{1993}).

\bibitem[{\citenamefont{Liepelt and Lipowsky}(2007)\citenamefont{Liepelt and Lipowsky}}]{Liepelt_Lipowsky2007}
  \bibinfo{author}{\bibfnamefont{S.} \bibnamefont{Liepelt}} \bibnamefont{and}
\bibinfo{author}{\bibfnamefont{R.}~\bibnamefont{Lipowsky}},
  \bibinfo{journal}{Phys.\ Rev.\ Lett.} \textbf{\bibinfo{volume}{98}},
  \bibinfo{pages}{258102} (\bibinfo{year}{2007}).

\bibitem[{\citenamefont{Wang et~al.}(2005)\citenamefont{Wang, Dou, and
  Wang}}]{Wang__Wang2005}
\bibinfo{author}{\bibfnamefont{H.}~\bibnamefont{Wang}},
  \bibinfo{author}{\bibfnamefont{S.-X.} \bibnamefont{Dou}}, \bibnamefont{and}
  \bibinfo{author}{\bibfnamefont{P.-Y.} \bibnamefont{Wang}},
  \bibinfo{journal}{Chin. Phys. Lett.} \textbf{\bibinfo{volume}{22}},
  \bibinfo{pages}{2980} (\bibinfo{year}{2005}).

\bibitem[{\citenamefont{Greulich et~al.}(2007)\citenamefont{Greulich, Garai,
  Nishinari, Schadschneider, and Chowdhury}}]{Greulich__Chowdhury2007}
\bibinfo{author}{\bibfnamefont{P.}~\bibnamefont{Greulich}},
  \bibinfo{author}{\bibfnamefont{A.}~\bibnamefont{Garai}},
  \bibinfo{author}{\bibfnamefont{K.}~\bibnamefont{Nishinari}},
  \bibinfo{author}{\bibfnamefont{A.}~\bibnamefont{Schadschneider}},
  \bibnamefont{and}
  \bibinfo{author}{\bibfnamefont{D.}~\bibnamefont{Chowdhury}},
  \bibinfo{journal}{Phys. Rev. E} \textbf{\bibinfo{volume}{75}},
  \bibinfo{pages}{041905} (\bibinfo{year}{2007}).

\bibitem[{\citenamefont{Leduc et~al.}(2004)\citenamefont{Leduc, Camp{\`a}s,
  Zeldovich, Roux, Jolimaitre, Bourel-Bonnet, Goud, Joanny, Bassereau, and
  Prost}}]{Leduc__Prost2004}
\bibinfo{author}{\bibfnamefont{C.}~\bibnamefont{Leduc}},
  \bibinfo{author}{\bibfnamefont{O.}~\bibnamefont{Camp{\`a}s}},
  \bibinfo{author}{\bibfnamefont{K.~B.} \bibnamefont{Zeldovich}},
  \bibinfo{author}{\bibfnamefont{A.}~\bibnamefont{Roux}},
  \bibinfo{author}{\bibfnamefont{P.}~\bibnamefont{Jolimaitre}},
  \bibinfo{author}{\bibfnamefont{L.}~\bibnamefont{Bourel-Bonnet}},
  \bibinfo{author}{\bibfnamefont{B.}~\bibnamefont{Goud}},
  \bibinfo{author}{\bibfnamefont{J.-F.} \bibnamefont{Joanny}},
  \bibinfo{author}{\bibfnamefont{P.}~\bibnamefont{Bassereau}},
  \bibnamefont{and} \bibinfo{author}{\bibfnamefont{J.}~\bibnamefont{Prost}},
  \bibinfo{journal}{Proc.\ Natl.\ Acad.\ Sci.\ USA}
  \textbf{\bibinfo{volume}{101}}, \bibinfo{pages}{17096}
  (\bibinfo{year}{2004}).

\bibitem[{\citenamefont{Konzack et~al.}(2005)\citenamefont{Konzack, Rischitor,
  Enke, and Fischer}}]{Konzack__Fischer2005}
\bibinfo{author}{\bibfnamefont{S.}~\bibnamefont{Konzack}},
  \bibinfo{author}{\bibfnamefont{P.~E.} \bibnamefont{Rischitor}},
  \bibinfo{author}{\bibfnamefont{C.}~\bibnamefont{Enke}}, \bibnamefont{and}
  \bibinfo{author}{\bibfnamefont{R.}~\bibnamefont{Fischer}},
  \bibinfo{journal}{Mol.\ Biol.\ Cell} \textbf{\bibinfo{volume}{16}},
  \bibinfo{pages}{497} (\bibinfo{year}{2005}).

\bibitem[{\citenamefont{Klumpp et~al.}(2005)\citenamefont{Klumpp,
  Nieuwenhuizen, and Lipowsky}}]{Klumpp__Lipowsky2005}
\bibinfo{author}{\bibfnamefont{S.}~\bibnamefont{Klumpp}},
  \bibinfo{author}{\bibfnamefont{T.~M.} \bibnamefont{Nieuwenhuizen}},
  \bibnamefont{and} \bibinfo{author}{\bibfnamefont{R.}~\bibnamefont{Lipowsky}},
  \bibinfo{journal}{Biophys.\ J.} \textbf{\bibinfo{volume}{88}},
  \bibinfo{pages}{3118} (\bibinfo{year}{2005}).

\bibitem[{\citenamefont{Seitz and Surrey}(2006)}]{Seitz_Surrey2006}
\bibinfo{author}{\bibfnamefont{A.}~\bibnamefont{Seitz}} \bibnamefont{and}
  \bibinfo{author}{\bibfnamefont{T.}~\bibnamefont{Surrey}},
  \bibinfo{journal}{EMBO J.} \textbf{\bibinfo{volume}{25}},
  \bibinfo{pages}{267} (\bibinfo{year}{2006}).

\bibitem{en1} In general, if motors may unbind in a conformation different
  from the one they had immediately after binding, one step per run
  corresponds to an incomplete chemical cycle. This incomplete cycle
  leads to a correction of the probabilities $P_i$ compared to the
  case, where the chemical cycle is completed in each step. This
  correction is thus of order 1/(number of steps per run) and has a
  very small effect for processive motors.

\bibitem{en2} The small non-zero value of $\epsilon_2$ in has been
  introduced to avoid an artificial absorbing state, given by a
  filament fully occupied by motors in state 2, from which the system
  cannot escape within the simplest models if $\epsilon_2=0$. An alternative would be to 
  introduce a reverse transition from state 2 to state 1 at the {\it same} position.
  
\bibitem{en3} The peculiar shape of the curve reflects the change in the dominant 
  unbinding pathway towards unbinding from state 2 at high unbound motor concentrations.

\bibitem[{\citenamefont{Lipowsky et~al.}(2006)\citenamefont{Lipowsky, Chai,
  Klumpp, Liepelt, and {M\"uller}}}]{ProcKanpur}
\bibinfo{author}{\bibfnamefont{R.}~\bibnamefont{Lipowsky}},
  \bibinfo{author}{\bibfnamefont{Y.}~\bibnamefont{Chai}},
  \bibinfo{author}{\bibfnamefont{S.}~\bibnamefont{Klumpp}},
  \bibinfo{author}{\bibfnamefont{S.}~\bibnamefont{Liepelt}}, \bibnamefont{and}
  \bibinfo{author}{\bibfnamefont{M.~J.~I.}~\bibnamefont{{M\"uller}}},
  \bibinfo{journal}{Physica A} \textbf{\bibinfo{volume}{372}},
  \bibinfo{pages}{34} (\bibinfo{year}{2006}).

\bibitem[{\citenamefont{Crevel et~al.}(2004)\citenamefont{Crevel, Nyitrai,
  Alonso, Weiss, Geeves, and Cross}}]{Crevel__Cross2004}
\bibinfo{author}{\bibfnamefont{I.~M.-T.~C.} \bibnamefont{Crevel}},
  \bibinfo{author}{\bibfnamefont{M.}~\bibnamefont{Nyitrai}},
  \bibinfo{author}{\bibfnamefont{M.~C.} \bibnamefont{Alonso}},
  \bibinfo{author}{\bibfnamefont{S.}~\bibnamefont{Weiss}},
  \bibinfo{author}{\bibfnamefont{M.~A.} \bibnamefont{Geeves}},
  \bibnamefont{and} \bibinfo{author}{\bibfnamefont{R.~A.} \bibnamefont{Cross}},
  \bibinfo{journal}{EMBO J.} \textbf{\bibinfo{volume}{23}}, \bibinfo{pages}{23}
  (\bibinfo{year}{2004}).

\bibitem[{\citenamefont{Frey and Vilfan}(2002)}]{Vilfan_Frey2002}
\bibinfo{author}{\bibfnamefont{E.}~\bibnamefont{Frey}} \bibnamefont{and}
  \bibinfo{author}{\bibfnamefont{A.}~\bibnamefont{Vilfan}},
  \bibinfo{journal}{Chem. Phys.} \textbf{\bibinfo{volume}{284}},
  \bibinfo{pages}{287} (\bibinfo{year}{2002}).

\bibitem[{\citenamefont{Carter and Cross}(2005)}]{Carter_Cross2005}
\bibinfo{author}{\bibfnamefont{N.~J.} \bibnamefont{Carter}} \bibnamefont{and}
  \bibinfo{author}{\bibfnamefont{R.~A.} \bibnamefont{Cross}},
  \bibinfo{journal}{Nature} \textbf{\bibinfo{volume}{435}},
  \bibinfo{pages}{308} (\bibinfo{year}{2005}).

\bibitem[{\citenamefont{Ross et~al.}(2006)\citenamefont{Ross, Wallace, Shuman,
  Goldman, and Holzbaur}}]{Ross__Holzbaur2006}
\bibinfo{author}{\bibfnamefont{J.~L.} \bibnamefont{Ross}},
  \bibinfo{author}{\bibfnamefont{K.}~\bibnamefont{Wallace}},
  \bibinfo{author}{\bibfnamefont{H.}~\bibnamefont{Shuman}},
  \bibinfo{author}{\bibfnamefont{Y.~E.} \bibnamefont{Goldman}},
  \bibnamefont{and} \bibinfo{author}{\bibfnamefont{E.~L.}
  \bibnamefont{Holzbaur}}, \bibinfo{journal}{Nature Cell Biol.}
  \textbf{\bibinfo{volume}{8}}, \bibinfo{pages}{562} (\bibinfo{year}{2006}).

\bibitem[{\citenamefont{Gennerich et~al.}(2007)\citenamefont{Gennerich, Carter,
  Reck-Peterson, and Vale}}]{Gennerich__Vale2007}
\bibinfo{author}{\bibfnamefont{A.}~\bibnamefont{Gennerich}},
  \bibinfo{author}{\bibfnamefont{A.~P.} \bibnamefont{Carter}},
  \bibinfo{author}{\bibfnamefont{S.~L.} \bibnamefont{Reck-Peterson}},
  \bibnamefont{and} \bibinfo{author}{\bibfnamefont{R.~D.} \bibnamefont{Vale}},
  \bibinfo{journal}{Cell} \textbf{\bibinfo{volume}{131}}, \bibinfo{pages}{952}
  (\bibinfo{year}{2007}).

\end{thebibliography}
\end{document}